\documentclass[9pt,leqno]{article}
\usepackage{color}
\usepackage{graphicx}

\newcommand{\be} {\begin{equation}}
\newcommand{\ee} {\end{equation}}
\begin{document}

\title{XTile: An Error-Correction Package for DNA Self-Assembly}
\author{Anshul Chaurasia , Sudhanshu Dwivedi,  Prateek Jain and Manish K. Gupta\\
Laboratory of Natural Information Processing,\\
Dhirubhai Ambani Institute of Information and Communication Technology,\\
Post Bag Number 4 , Near Indroda Circle,
Gandhinagar, Gujarat   382007,
India\\
Email: \{anshul\_chaurasia, sudhanshu\_dwivedi, prateek\_jain\_2006\}@daiict.ac.in,\\ m.k.gupta@ieee.org 
}
\maketitle

\vspace*{0.5cm}
%\vspace*{1cm}

\begin{abstract}
Self-assembly is a process by which supramolecular species form spontaneously from their components. This process is ubiquitous throughout the life chemistry and is central to biological information processing. It has been predicted that in future self-assembly will become an important engineering discipline by combining the fields of bio-molecular computation, nano-technology and medicine. However  error-control is a key challenge in realizing the potential of self-assembly. Recently many authors have proposed several combinatorial error correction schemes to control errors which have a close analogy with the coding theory such as Winfree's proofreading scheme and its generalizations by Chen and Goel and compact scheme of Reif, Sahu and Yin. In this work, we present an error-correction computational tool XTile that can be used to create input files to the Xgrow simulator of Winfree by providing the design logic of the tiles and it also allows the user to apply proofreading, snake and compact error correction schemes.
\end{abstract}
%\vspace{1.5cm}
\vspace*{0.5cm}
%%%%%%%%%%%%%%%%%%%%%%%%%%%%%%%%%%%%%%%%%


\vspace{0.5cm}

\newpage

%%%%%%%%%%%%%%%%%
\section{Introduction}
Self-assembly is a natural phenomenon observed at many places in nature such as formation of galaxies, formation of coral reefs,  crystal growth etc. In $1996$ Erik Winfree of California Institute of Technology, showed that it can be used to perform nano-scale computations. This paved way for the birth of algorithmic self-assembly utilizing knowledge of three fields - DNA Nanotechnology~\cite{byklwrs00} (due to the pioneering work of Ned Seeman in $1980s$), DNA Computing (due to the pioneering work of L. Adleman in $1994$) and Tiling Theory ~\cite{wang63} (due to pioneering work of H. Wang who showed that zigsaw shaped colored tiles can simulate universal Turing machine). Winfree formulated the idea of molecular Wang tile using all this and showed that it can simulate universal Turing machine using abstract tile assembly model (aTAM) ~\cite{win98}.  As any information processing is prone to errors when we perform any experiment in wet lab using DNA self assembly many errors are bound to occur. Using the idea of redundancy many error correction schemes have been proposed and of them, proofread, snake and compact error correction schemes have been studied well ~\cite{rsy04,wibe04,chgo04}.  Winfree's tile assembly model can be simulated using a program developed by him called Xgrow ~\cite{xgrow98}.  The input to Xgrow is  a .tiles file where each normal tile is represented by a set of  $4$ glue numbers (representing the four directions) and it also has a line specifying the respective glue strength. Many such standard  .tiles files are distributed with the Xgrow package. 
In this work, we present an error correction package XTile that can be used to create such input files for the Xgrow simulator of Winfree by providing the design logic of the tiles and it also allows the user to apply proofreading~\cite{wibe04}, snake~\cite{chgo04} and compact~\cite{rsy04} error correction schemes.  For further details on algorithmic self-assembly and various error-correction schemes the reader is referred to excellent papers ~\cite{wibe04, rsy04, chgo04} and thesis ~\cite{win98}.
This paper is organized as follows. In section $2$, we give a brief overview of the error correction schemes and section $3$ provides graphical user interface overview. In section $4$, we describe  major components of the tool.  Section $5$ provides the link from where one can download the tool. 
%%%%%%%%%%
\section {Error-Correction Schemes}
The main source for errors is the unfavorable attachment of tiles. Out of many possible errors three have been studied well. The first one is called {\bf mismatch error}. This is an unfavorable attachment that only partially matches the adjacent tiles and can occasionally become locked into place by succeeding attachments. This can be corrected by proofreading, snake and compact error correction schemes. The proofread error-correction scheme was given by  Winfree et.al.~\cite{wibe04}. In proofread a normal tile is broken into  $ m \times m $ tiles with the original glues at the outer side of the outer tiles. The inner tiles have new and distinct glues around them. This gives redundancy and reduces the error. The second type of error is called {\bf facet error}. In this a title attaches unfavorably to a facet and in turn allows the attachment of incorrect tiles nearby. This error is reduced by proofread error correction. The third type of error is called {\bf nucleation error} which occurs when an assembly grows from a tile other than the designated seed tile. To prevent this one requires Zig-zag tiles. In snake error correction scheme, a normal tile is broken into  $m \times m (m = 2n \;\mbox{for snake scheme})$ parts (with the original glues on the side tiles) according to the rules given by Chen et. al. ~\cite{chgo04}. This scheme is similar to proofread but its tile growing pattern differs and makes it better. This scheme helps to prevent mismatch and nucleation error. The other remarkable error correction scheme is given by Reif et. al.~\cite{rsy04} known as compact error correction scheme which was later generalized to $k$-way compact.  In this scheme we introduce redundancy in  the bits around the sides and use new rules to do the computation. Currently we have implemented the compact  $2$-way and $3$-way scheme. 
%%%%%%
\section{Graphical User Interface Overview}
The graphical user interface (GUI) for XTile has been developed to act as an interface between the background computational processing and the various I/O operations that are needed to be performed through the participation of the user. We have created the java applet in $3$ parts. The XTile $1.0$ is a basic .tile file generator, XTile $1.1$ does proofread ($m \times m$) and snake ($2n \times 2n$) including the basic tile generation and finally XTile $1.2$ creates compact $2$-way and $3$-way tiles files. 

\subsection{ Basic Tile Design}
The XTile  accepts basic boolean logic inputs from east and south and the corresponding boolean functions at north and west as outputs as described by Winfree and Reif et al \cite{wibe04,rsy04}. The user can enter any number of inputs/outputs  by separating commas at each side. The background processing takes care of providing to the user the reinterpretation of the boolean logic of inputs/outputs on the four faces in the form of glues (denoted by different numbers). The user can then choose from a variety of glues that have been previously used, taking into account the binary code which the glue stands for (this being displayed in a tabular format for the user to choose from while operating the XTile) or even add more glue types by utilizing sequence number that are still unused (as indicated in the tabular format that displays glue numbers in correspondence with the binary code they represent). Then by entering the glue numbers for frame tiles the user can actually input the binary code for the frame tiles which essentially supply inputs to the computational tiles. To represent an absence of any glue the glue number '0' is used. Thus by using the inputs for the glues for the frame tiles and the design of the computational tile (in terms of boolean logic) a '.tiles' file is generated.The above functionality had been implemented in XTile version $1.0.$

\subsection{Proofread and Snake Error Correction Schemes Tile Design}
On the .tiles files  obtained in previous section two error correction schemes, namely proofread and snake can be applied resulting in the generation of the desired error correction .tiles file.  For this the GUI interface requires the user to enter the value of $m$, an integer, so as to apply a $m \times m$ proofread  error correction scheme or a $m \times m$ snake error correction scheme (in this case $m$ is even). After accepting an integer value for $m$, the background processing takes place and the .tiles file with the respective error correction scheme applied is generated. The above functionality had been implemented in XTile version $1.1$ along with the functionality of generation of basic .tiles files, given the tile design, which was provided since XTile version $1.0$ onwards. 

\subsection{Compact Error Correction Schemes Tile Design}
The generation of a .tiles file with the compact error correction scheme applied on it varies from the above mention steps that were followed in case of application of proofread or snake error correction scheme. Here the user has to provide the tile design on which the compact error correction scheme is desired to be applied. There can be any number of inputs/outputs on a particular face and the rest of the inputs/outputs on other faces may or may not be dependent on the values of the other inputs/outputs. After confirming the tile design by enter the inputs/outputs on the four faces (North, East, South, West) of the tile, the user needs to input the value of inputs for horizontal and vertical tiles which would be applied to the given tile design following the rules of compact error correction scheme. The above functionality had been implemented in XTile version $1.2$. 

%%%%%%%%
\section{Detailed Description of GUI}
Developed as Java based applets, the GUI primarily serves to facilitate the user to obtain desired .tiles files after he provides the tile design and allows for application of various error correction schemes as desired, provided the required inputs are supplied as per the needs of the respective error correction schemes.

\subsection{The XTile $1.0$ has following six major components:}
\textbf{Inputs/outputs windows} are used to accept the boolean logic based inputs/outputs for each of the four faces (North, East, South, West) of the tile. Multiple inputs/outputs can be given for a particular face by separating each boolean expression from others by means of commas. After filling in the values of inputs/outputs as boolean expressions in all the four windows the user needs to click on the submit button. \\
%\item
\textbf{Confirm button} needs to be pressed after the user gets to view what he/she filled in the inputs/outputs windows appeared on the applet screen (after pressing the submit inputs button). This ensures that user has double checked what he had entered before the inputs/outputs are sent to background processing to generate glue tables in the glue table window.\\
%\item 
\textbf{Glue table window} displays glues that have been used till now by the tile design as given by the user (by processing the boolean logic based inputs/outputs) in correspondence to the binary code they stand for. So by looking at the glue table displayed in the glue window user can choose from a variety of glues that have been previously used or add more glue types by utilizing sequence number that are still unused.\\
%\item 
\textbf{Frame glue window} needs to fill by the user with the sets of glues for as many frame tiles the user wants. The glues need to be denoted by integer numbers (which can either be chosen from the glue table as displayed in the glue table window or even be a user given number). The user needs to decide as to what binary digits he needs to send on each of the faces of a frame tile. While deciding this, the user should keep in mind as to how many individual boolean expressions are there on a particular face. This will decide the number of bits in the binary code that the user needs to enter, by choosing the corresponding glue numbers for that particular binary code. So a set of four numbers (one for each of the four faces of a tile) need to be given for specifying a frame tile. Multiple frame tiles can be given but they need to be separated from each other by semicolons. No terminating semicolon is needed at the end of the last frame tile specified. \\
%\item 
\textbf{Frame glue button} needs to be pressed after the user enters the glue numbers (in sets of four) for as many frame tiles as he wishes to create, so that background processing can follow.\\
%\item 
\textbf{.tiles file display window} displays the final .tiles file as desired by the user to be generated according to the tile design he wishes to implement.
%\end{itemize}

\subsection{The XTile $1.1$ has following additional four components as compared to XTile $1.0$:}
%\begin{itemize}
%\item
 \textbf{Proofread input window} accepts from the user the value of $m$, an integer, so as to apply a $m \times m$ proofread error correction scheme.\\
%\item 
\textbf{.tiles with proofread error correction applied window} displays the .tiles file with the proofread error correction scheme applied which is generated after accepting an integer value for $m$.\\
%\item 
\textbf{Snake input window} accepts from the user the value of $m$, an even integer, so as to apply a $2n \times 2n$ snake error correction scheme.\\
%\item 
\textbf{.tiles with snake error correction applied window} displays the .tiles file with the snake error correction scheme applied which is generated after accepting an integer value for $m$.
%\end{itemize}

\subsection{The XTile $1.2$ has following five major components:}

%\begin{itemize}
%\item 
\textbf{Inputs/outputs windows} are used to accept the boolean logic based inputs/outputs for each of the four faces (North, East, South, West) of the tile. Multiple inputs/outputs can be given for a particular face by separating each boolean expression from others by means of commas. After filling in the values of inputs/outputs as boolean expressions in all the four windows the user needs to click on the submit button. \\
%\item 
\textbf{Confirm button} needs to be pressed after the user gets to view what he/she filled in the inputs/outputs windows appeared on the applet screen (after pressing the submit inputs button). This ensures that user has double checked what he had entered before the inputs/outputs are sent to background processing to generate glue tables in the glue table window.\\
%\item 
\textbf{Frame glue windows} accepts from the user the binary values for the vertical and horizontal frame tiles which are processed according to the rules of the compact error correction scheme.\\
%\item 
\textbf{Frame glue button needs} to be pressed after the user enters the glue numbers for vertical and horizontal frame tiles as he wishes to create, so that background processing can follow.\\
%\item 
\textbf{.tiles with compact error correction applied window} displays the final .tiles file as desired by the user to be generated according to the tile design and the compact  error correction scheme he wished to implement on it.
%\end{itemize}
%%%%
\section{Software Availability}
The tool XTile is available for download at http://www.guptalab.org/xtile. The user manual and online version of the tool is also available on the webpage. 
%\bigskip
%%%%%
\section{Acknowledgements.}
The authors would like to thank Archit Jain for useful discussions.
%%%%%%%%%%%%%%%%%%%%  references  %%%%%%%%%%%%%%%%%%%%%%
\noindent
%%%%%%%%%%%%%%%%%%%%%%%%%

\end{document}